\documentclass[pre,aps,twocolumn,showpacs]{revtex4}
\usepackage{amssymb}
\usepackage{amsmath}
\usepackage{epsfig}
\usepackage{mathrsfs}
\usepackage{sidecap}
\begin{document}

\title
{\bf Square Lattice Gases with Two- and Three-body\\
Interactions Revisited\\}

\author{Junqi Yin and D. P. Landau}

\affiliation{Center for Simulational Physics, University of
Georgia, Athens, Georgia 30602, USA\\
}

\date{\today}

\begin{abstract}
Monte Carlo simulations have been used to study the phase diagrams
for square Ising-lattice gas models with two-body and three-body
interactions for values of interaction parameters in a range that
has not been previously considered. We find unexpected qualitative
differences as compared with predictions made on general grounds.
\end{abstract}

\pacs{68.35.Rh, 64.60.Cn, 05.70.Jk, 64.60.Kw}
\maketitle

\section{Introduction}
\begin{SCfigure*}
\includegraphics[width=0.75\columnwidth]{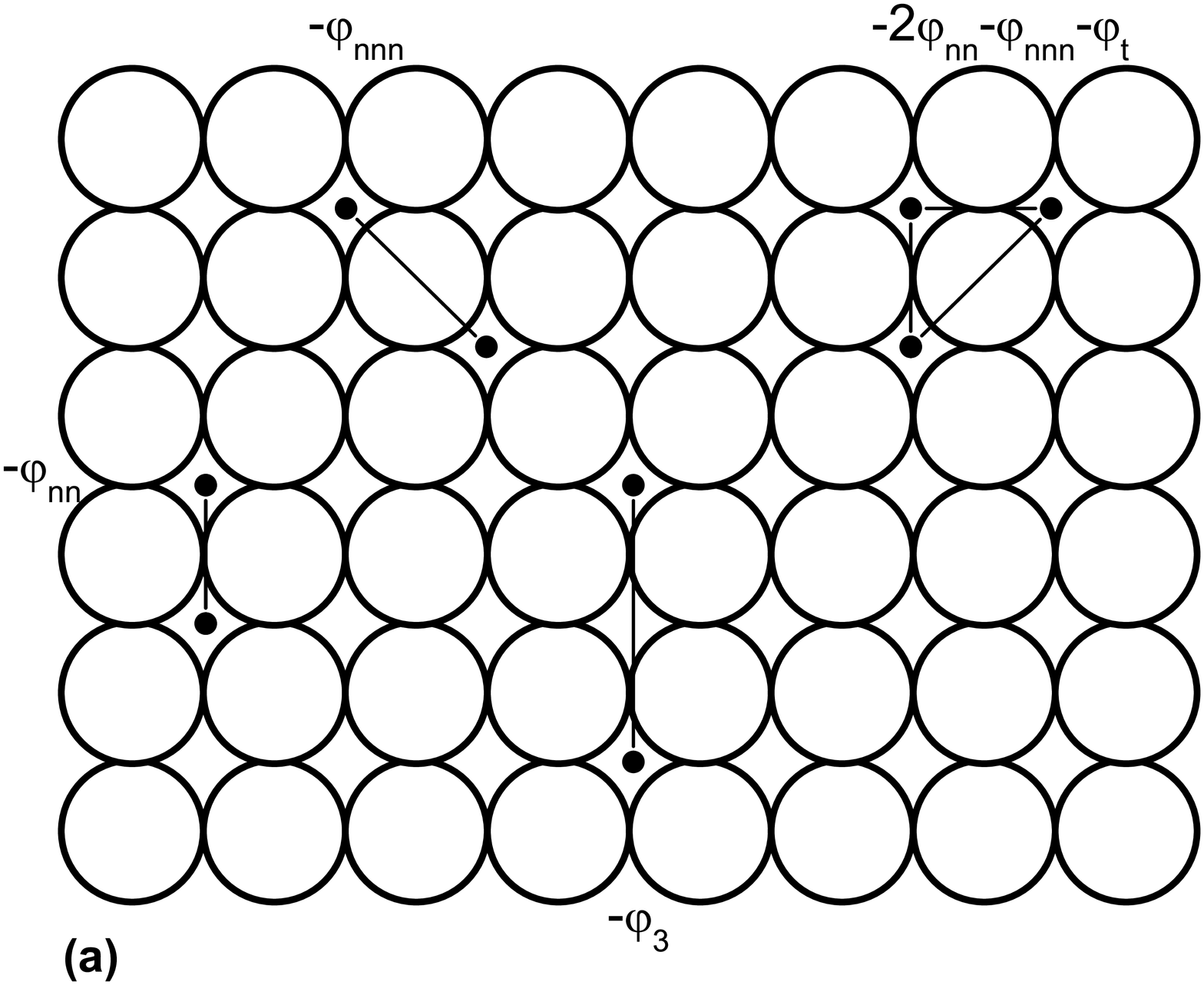}
\hspace{0.1cm}
\includegraphics[width=0.75\columnwidth]{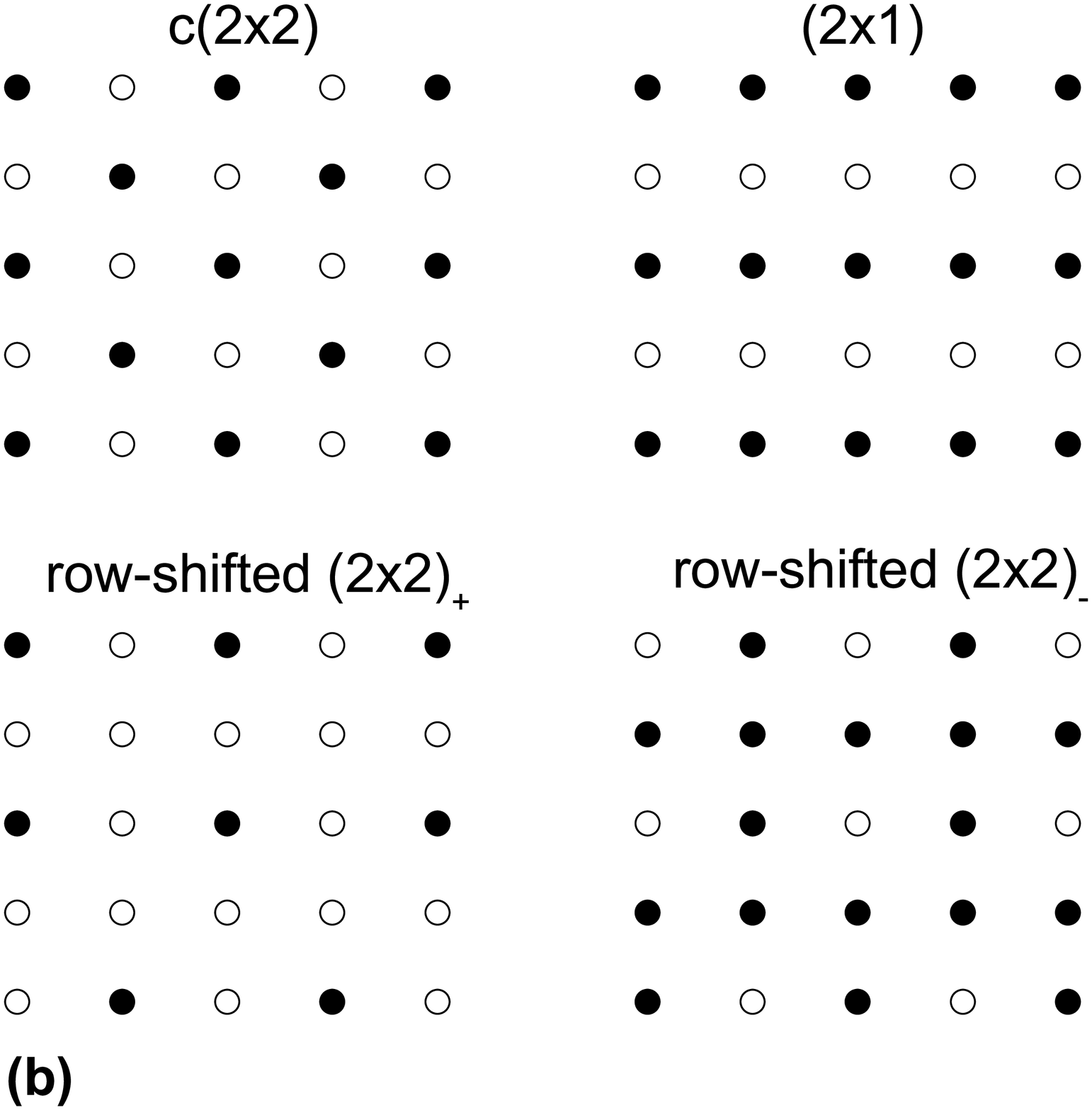}
\caption{(a) Schematic view of the (100) surface of a substrate whose periodic
potential provides a square lattice of preferred adsorption sites.  The
interactions used in this study are shown schematically as straight lines
between adatoms which are represented by the filled circles.  (b) Unit
cells of the ordered overlayer structures discussed in the text.} \label{f1}
\end{SCfigure*}

Experimental studies of phase transitions in adsorbed monolayers
have resulted in examination of order-disorder transitions in
lattice gas Ising models which represent the occupation of the
periodic minima in the substrate potential\cite{das78,bin89}.  Such
models, usually containing two or more competing two-body
interactions, have been studied by Monte Carlo
simulations\cite{bin80,bin82} which have determined the location and
nature of the resultant phase boundaries.  Typical ordered phases
which are found for square lattice models with near-neighbor
coupling are shown in Fig.~\ref{f1} along with low density and high
density disordered states termed lattice gas (L.G.) and lattice
liquid (L.L.), respectively.
 Experimentally observed asymmetries in phase
boundaries as a function of coverage can be explained by lattice gas
models if three-body interactions are introduced as well. Theoretical studies of adatom-adatom interactions find surface mediated many-body couplings\cite{ein73}.  A square
lattice model with first- and second-neighbor two-body coupling and
weak three-body interactions was investigated\cite{bin81} in an
attempt to clarify the phase transitions for H on  Pd(001)  and
predictions were made for the case of even larger three-body
coupling. Monte Carlo simulations showed that the inclusion of
three-body interactions did make the transition asymmetric about 50
percent coverage and could even force the tricritical point on one
side of the phase boundary to zero temperature.  Such an asymmetry
can also be interpreted in another form of non-additive
interactions\cite{mil81}, and a recent Monte Carlo study shows the
effect of various cases of strength of interactions\cite{pin08}.

In this paper, we present the results of an investigation of the
model proposed in Ref \cite{bin81} with moderate to large three-body
interactions using Monte Carlo simulations.  In the next section, we
shall review some appropriate background and in Sec. III we present
our results for two different values of interaction parameters which
yield qualitatively different phase diagrams.  We conclude in Sec.
IV.

\section{Background}

\begin{figure*}
\includegraphics[width=0.7\columnwidth]{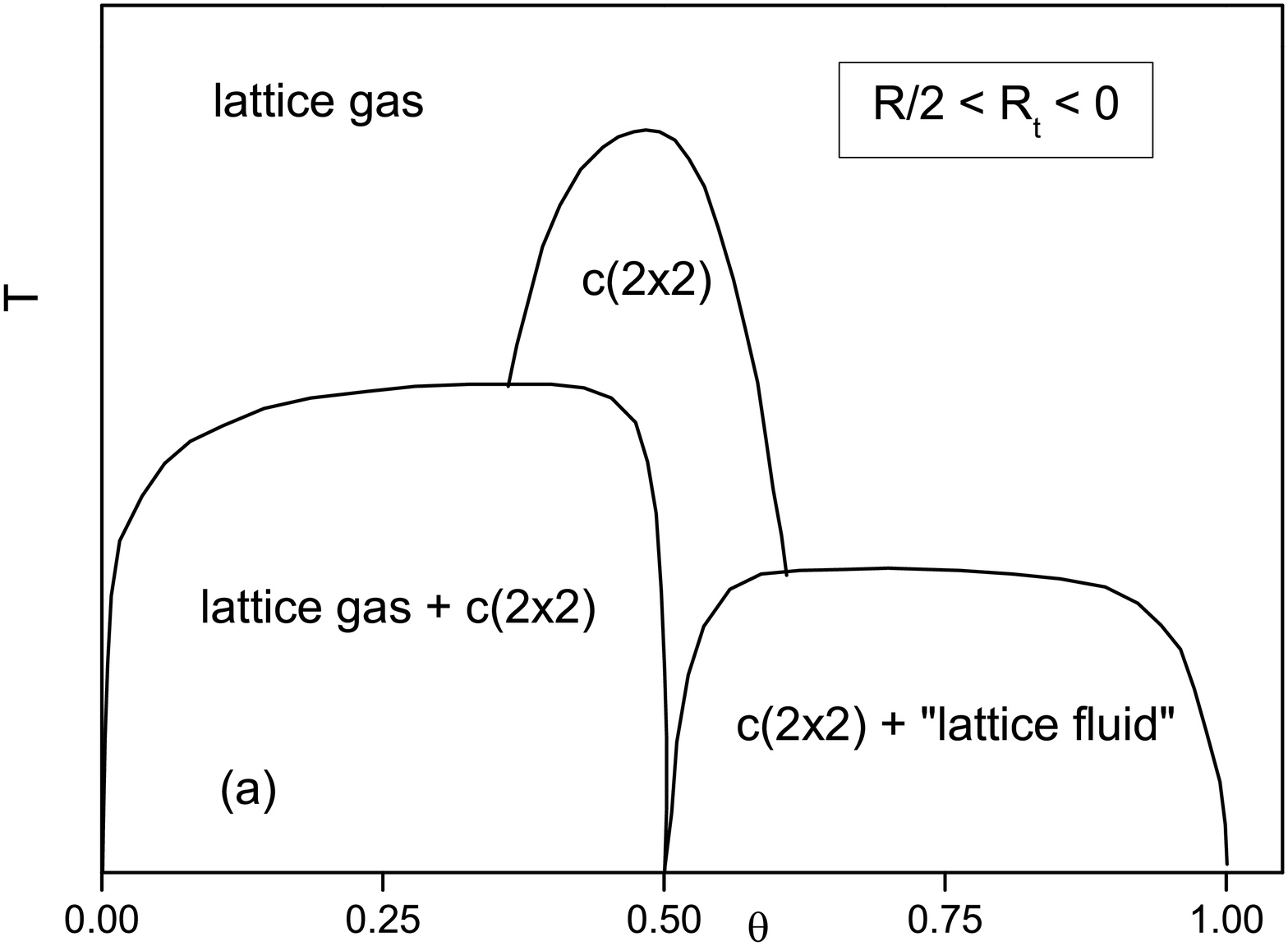}
\includegraphics[width=0.7\columnwidth]{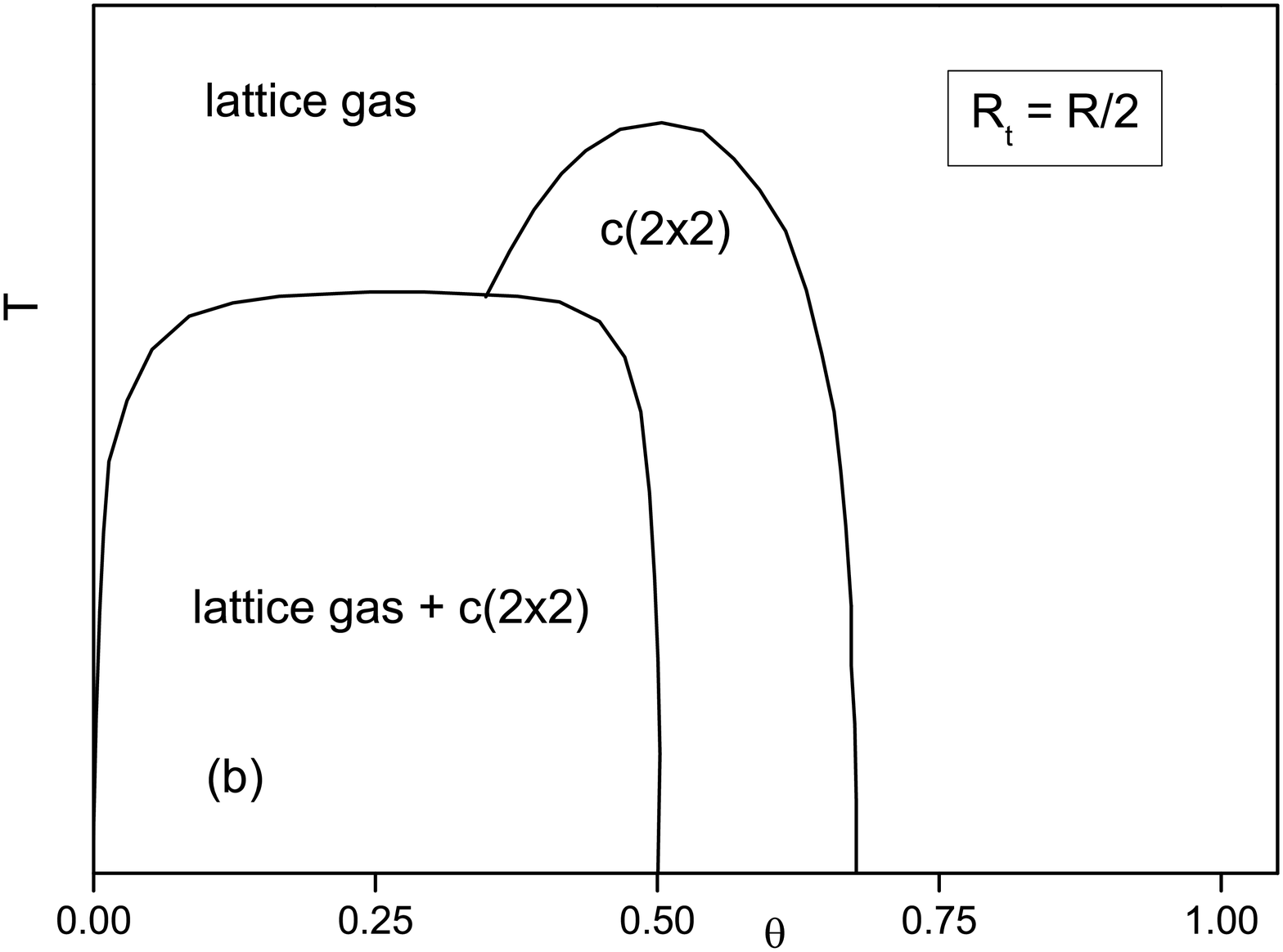}
\includegraphics[width=0.7\columnwidth]{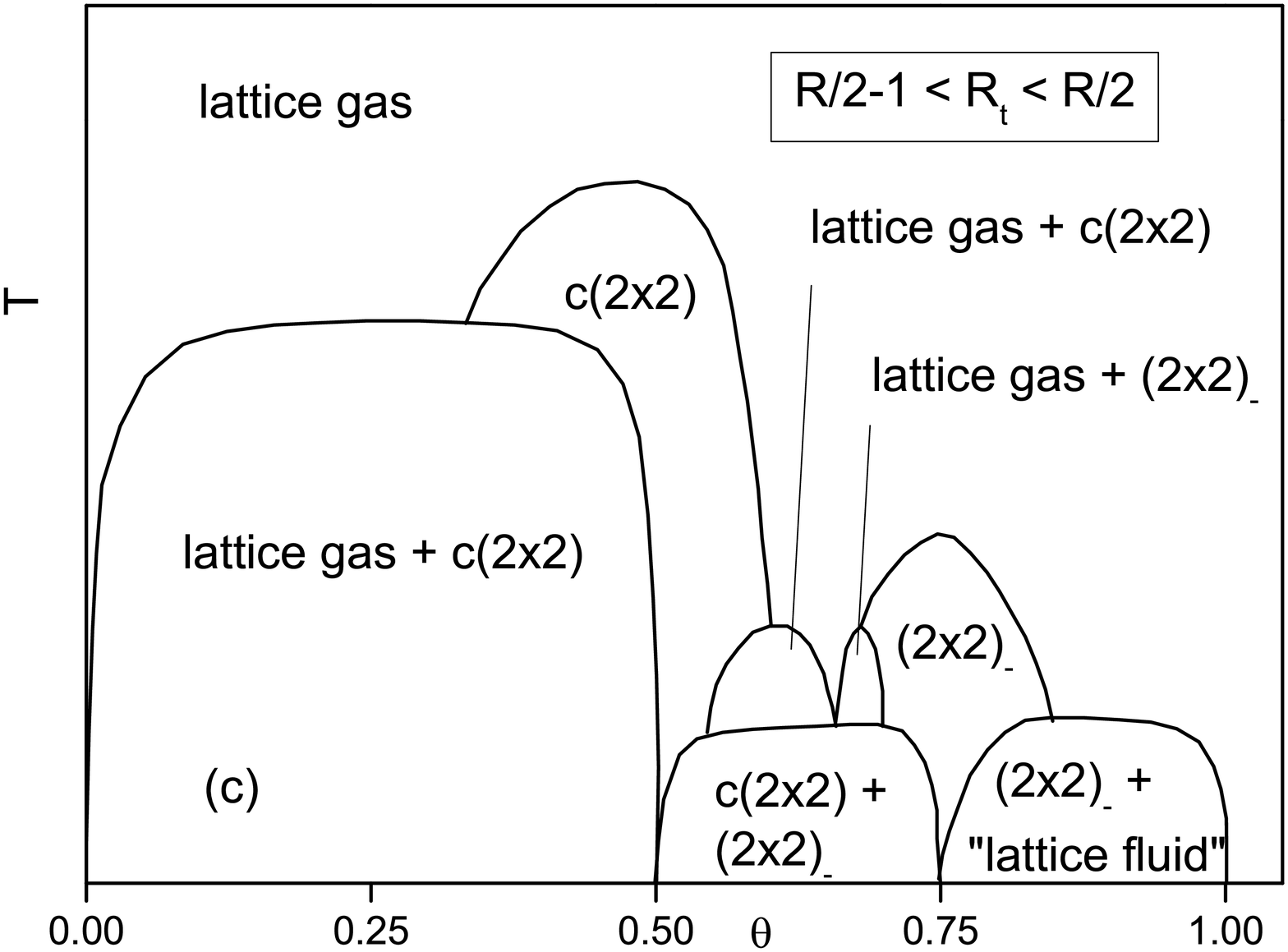}
\includegraphics[width=0.7\columnwidth]{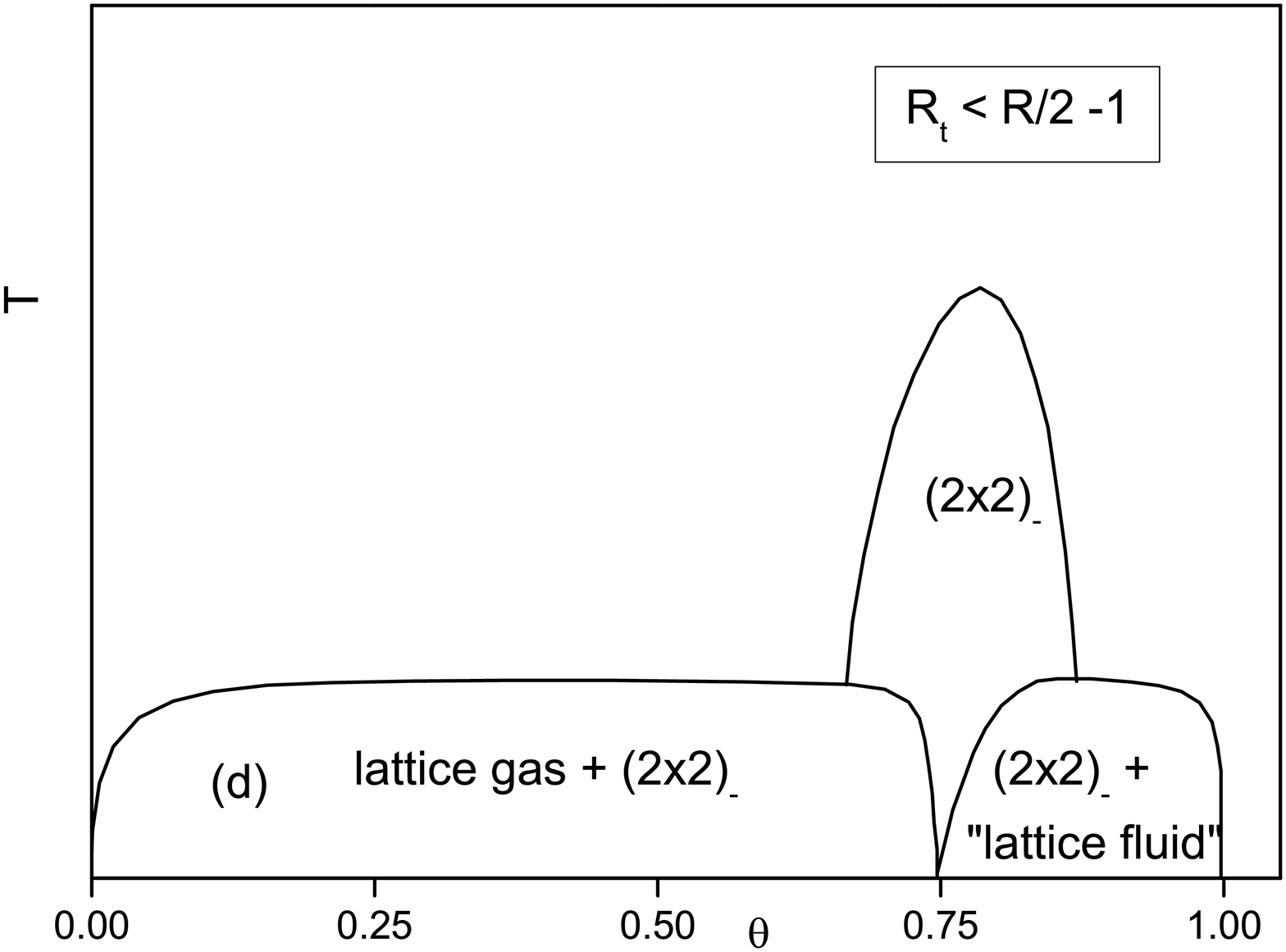}
\caption{Possible schematic temperature-coverage phase diagrams for various
choices of $R_t$ (from Ref. ~\cite{bin81}).} \label{f2}
\end{figure*}

A lattice gas model is a collection of atoms whose positions may take on only
discrete positions which form a periodic array, in this case a simple square
lattice.  A configuration of this lattice is defined by site occupation
variables $c_i$ where $c_i = 1$ if site $i$ is occupied and $c_i = 0$ if the
site is empty.  The Hamiltonian which we use includes interaction $\varphi_{nn}$
between nearest-neighbors, $\varphi_{nnn}$ between next-nearest-neighbors, and
$\varphi_t$ between neighbors on a triangle inscribed within a square made up of
nearest-neighbors:

\begin{eqnarray}
{\cal H} - \mu N_a =&-& (\epsilon + \mu) \sum_i c_i - \varphi_{nn} \sum_{i \neq j} c_i c_j \nonumber \\
                    &-& \varphi_{nnn} \sum_{i \neq k} c_i c_k - \varphi_t \sum_{i \neq j \neq k} c_i c_j c_k,
\end{eqnarray}
and the coverage of the lattice is given by

\begin{eqnarray}
\theta = \sum_i c_i
\end{eqnarray}
This model may be transcribed to the Ising model by the
transformation to spin variables $\sigma = 1 - 2c_i$, thus giving
rise to the Hamiltonian

\begin{eqnarray}
{\cal H} = &-&H\sum_i \sigma_i - J_{nn}\sum_{i \neq j} \sigma_i \sigma_j \nonumber \\
           &-&J_{nnn} \sum_{i \neq k} \sigma_i \sigma_k -J_t \sum_{i \neq j \neq k} \sigma_i \sigma_j \sigma_k,
\end{eqnarray}
where the Ising model interaction parameters are related to the
lattice gas couplings by

\begin{eqnarray}
J_{nn} = \frac{1}{4} \varphi_{nn} + \frac{1}{2} \varphi_t
\end{eqnarray}
\begin{eqnarray}
J_{nnn} = \frac{1}{4} \varphi_{nnn} + \frac{1}{4} \varphi_t
\end{eqnarray}
\begin{eqnarray}
J_t = -\frac{1}{8} \varphi_t
\end{eqnarray}
\begin{eqnarray}
H = {-\frac{1}{2}} (\epsilon+\mu)-\varphi_{nn}-\varphi_{nnn} - \frac{3}{2}
\varphi_t
\end{eqnarray}
The magnetization is then related trivially to the coverage

\begin{eqnarray}
m = 1-2\theta.
\end{eqnarray}
Because of technical considerations, it is generally easier to carry
out simulations in the magnetic (Ising) representation, and
additional symmetries
 often become obvious in this approach.  For example, in the Ising
 representation it is easy to see that the phase diagram must be
 symmetric in the absence of three-body interactions.  Throughout the
 remainder of this paper we shall normalize all quantities by the
 nearest-neighbor coupling $J_{nn}$ and define $R=J_{nnn}/J_{nn}$
 and $R_t=J_t/J_{nn}$. As a function of the relative strength of the three-body interactions, Binder
 and Landau suggested four different schematic phase diagrams (shown in
 Fig.~\ref{f2}) to describe the range of possible behavior due to the inclusion
 of three-body coupling, and using Monte Carlo simulations verified the
 nature of the phase diagrams with weak three-spin coupling shown in the
 upper portion of Fig.~\ref{f2}.
 The remaining two diagrams were not based on any
 explicit calculation but were guessed as extensions of what was then
 believed to be the field-dependent behavior in the absence of three-spin
 interactions.  Our views of the correct behavior in this latter case
 have changed, and given the complexity of phase diagrams in other
 two-dimensional systems with competing interactions, the predictions should
 be regarded with care.

\begin{figure}
\includegraphics[width=0.8\columnwidth]{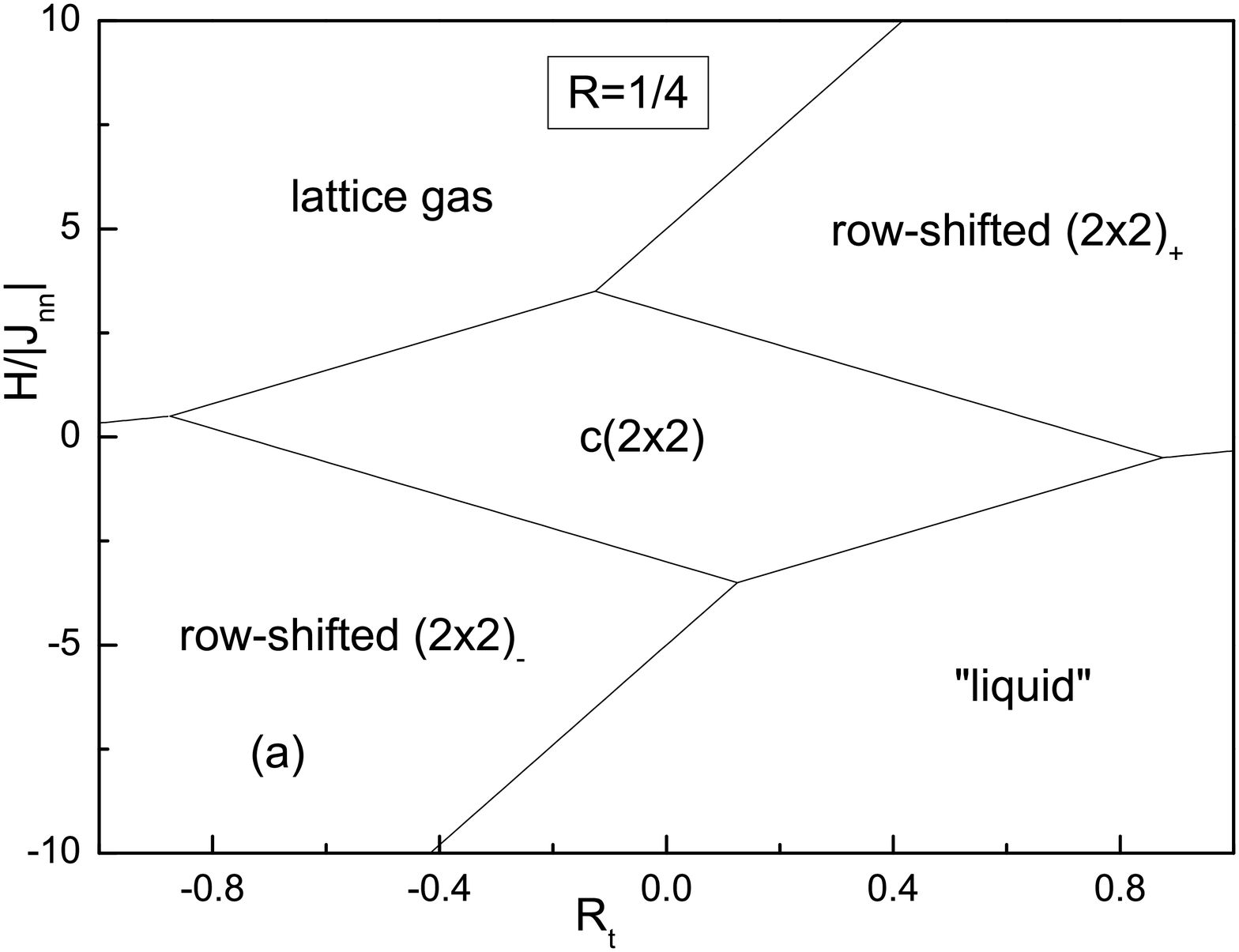}
\includegraphics[width=0.8\columnwidth]{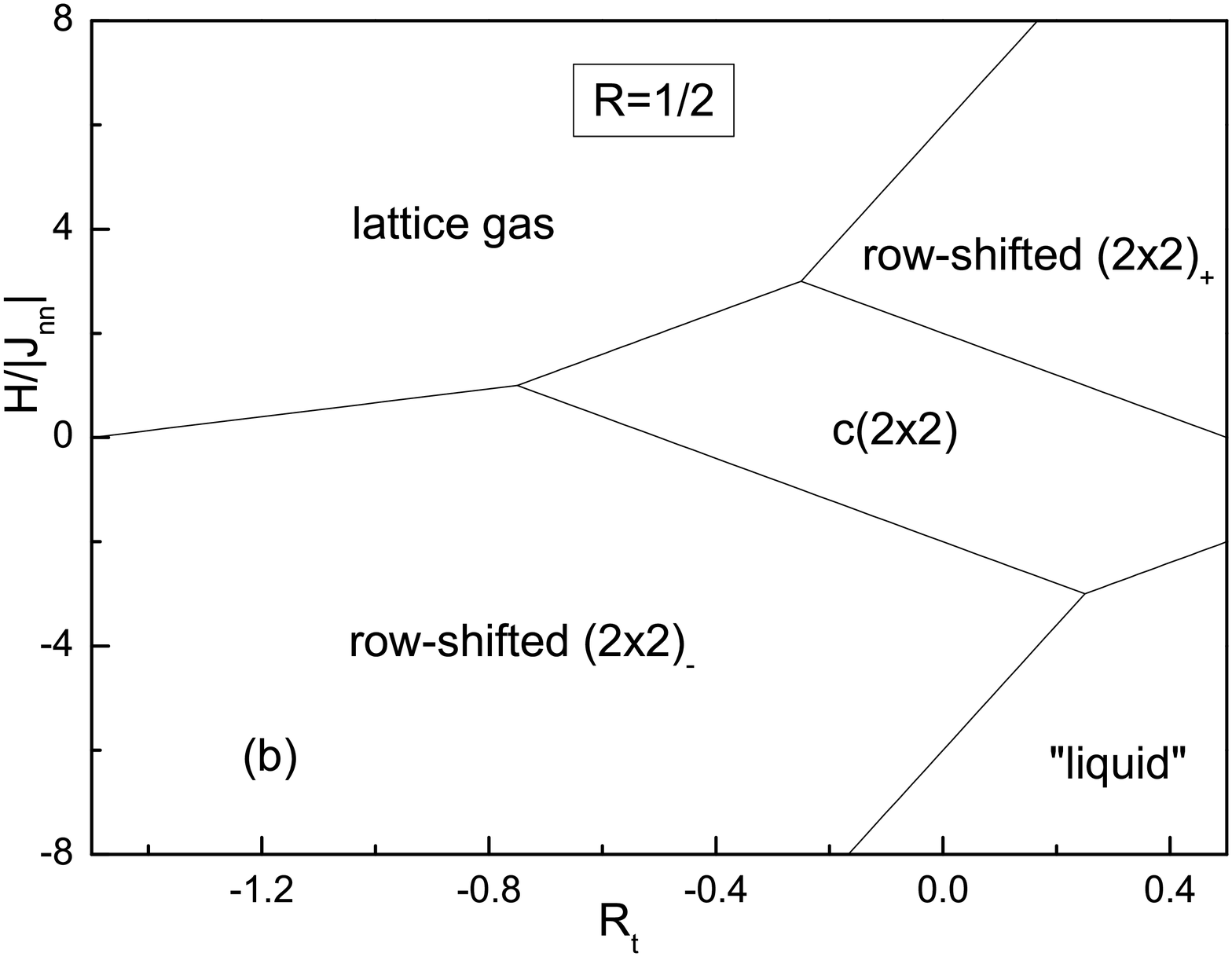}
\caption{Groundstate phase diagrams for the Ising square lattice with pairwise and
three-spin interactions:  (a) $R = 1/4$; (b) $R = 1/2$.} \label{f3}
\end{figure}

At $T=0$ the energies of each ordered state as well as the lattice
gas and lattice liquid phases may be calculated without difficulty
and the intersections of the energy vs. chemical potential lines
locate the transitions between neighboring phases.  Of course, this
is also valid in the Ising representation, and in Fig.~\ref{f3} we
show the ground state phase diagrams as a function of $R_t$ and $H$
which we obtain for two different values of $R$ which are in the
parameter region discussed by Binder and Landau\cite{bin81} but not
actually simulated.

One interesting feature of this diagram is that the $(2\times2)$
state is actually degenerate in that either alternate rows or
alternate columns may be shifted randomly by one lattice constant
without any cost in energy\cite{bin80,yin09}. This ``row-shifted
$(2\times2)$'' state has been seen before and the nature of the
finite temperature transition to the disordered state is a matter of
some disagreement. We believe that as long as the ground states do
not change, the specific choice of parameters is not important and
we have simply chosen values for which the $c(2\times2)$ and
row-shifted $(2\times2)$ phases are stable over relatively large
ranges of fields in Fig.~\ref{f3} at which multiple phases become
degenerate.

We have used the parallel tempering algorithm with GPU accelerated
techniques\cite{yin09} to study the thermodynamic properties of this
model from which phase diagrams can be deduced. Spins were placed on
$L\times L$ square lattices with periodic boundary conditions and
were flipped using a Metropolis transition probability. Typically,
$10^6$ to $10^7$ Monte Carlo steps are used to collect data for each
run and 3 to 6 independent runs are taken to calculate standard
statistical error bars, and in all the plots of data and analysis
shown in following sections, if error bars are not shown they are
always smaller than the size of the symbols. Lattice sizes from
$L=32$ to $L=300$ were simulated and the data were interpreted
within the context of finite size scaling\cite{pri90}. Most of the simulations
were carried out on a GeForce GTX285 graphics unit. In addition to
internal energy, specific heat, and magnetization, we calculated
order parameters, e.g., $m_{c(2\times2)}, m_{2\times1}$, etc., for
the various ordered states shown in Fig.~\ref{f1}, and 4th order
cumulant $U$ is defined in terms of the order parameter accordingly.

\section{Results}

A. ~~ $R=1/4, R_t=-1/4$

Bulk properties such as the specific heat peak, temperature
dependence of the 4th order cumulant of the order parameter, etc.,
were used to determine the location of phase transitions.
\begin{figure}
\includegraphics[width=0.96\columnwidth]{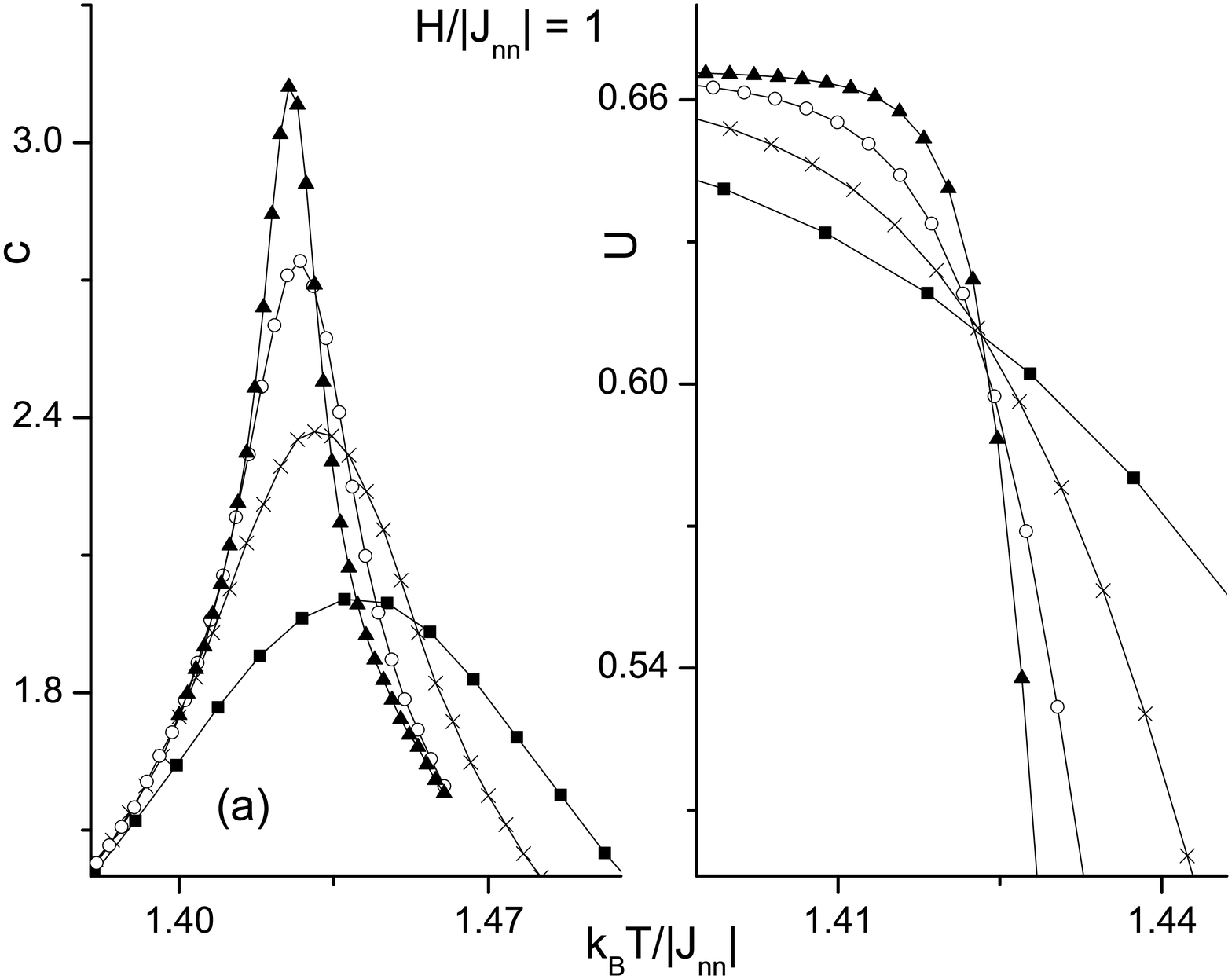}
\vspace{0.2cm}\\
\includegraphics[width=0.96\columnwidth]{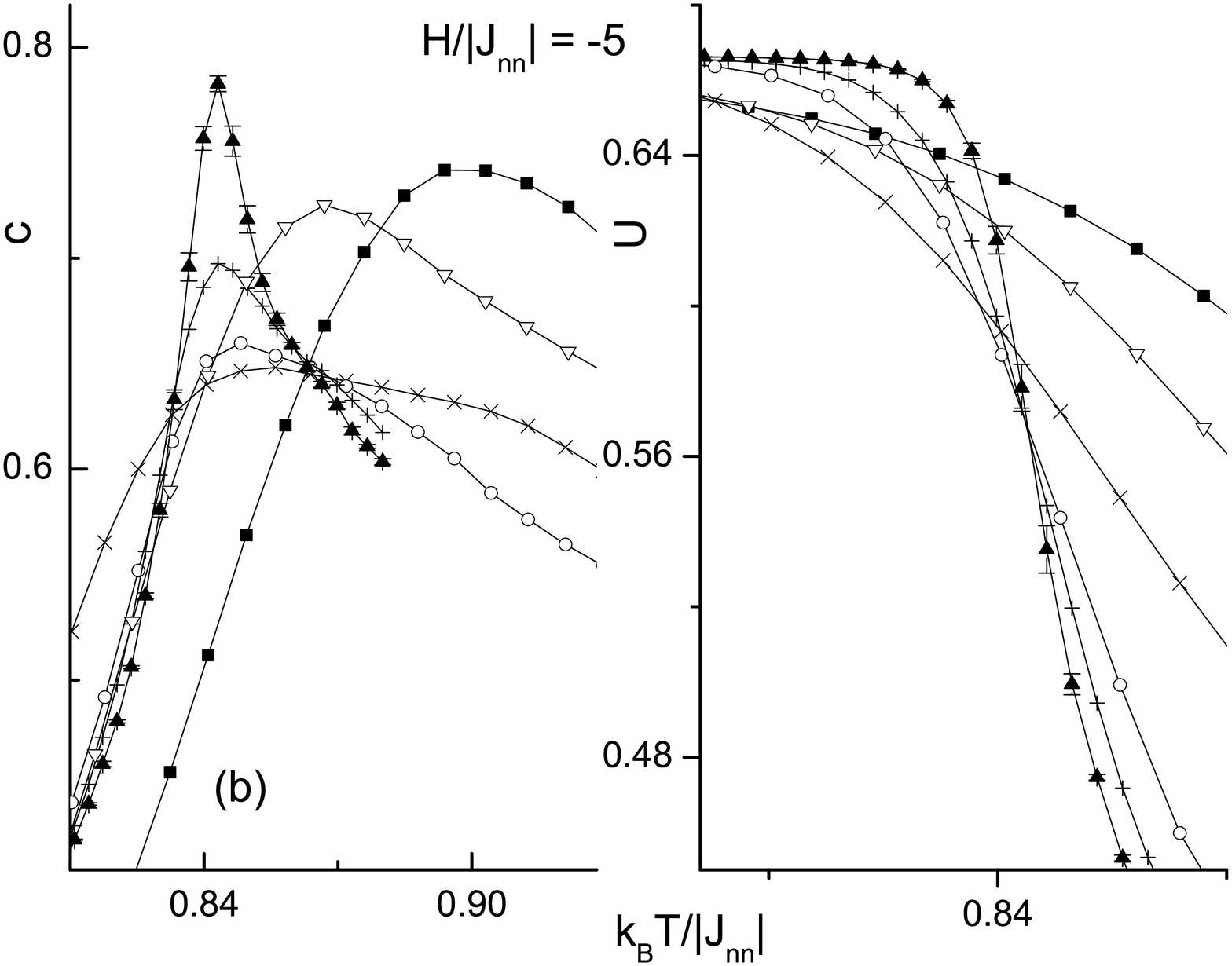}
\caption{Bulk properties for $R = 1/4, R_t = -1/4$: Specific heat
$c$ and 4th order cumulant $U$ of the corresponding order parameter
for (a) $H/|J_{nn}| = 1$. Data are for: L=32, $\blacksquare$; L=64, $\times$; L=128,
$\circ$; L=256, $\blacktriangle$. (b) $H/|J_{nn}| = -5$. Data are for L=30, $\blacksquare$;
L=40, $\triangledown$; L=64, $\times$; L=128, $\circ$; L=168, $+$; L=256, $\blacktriangle$.} \label{f4}
\end{figure}
Sample data for the 4th order cumulant and specific heat are shown
in Fig.~\ref{f4}; the specific heat peaks diverge with increasing
lattice sizes for fields above $H/|J_{nn}| = -2.0$, but for fields
more negative and close to the $c(2\times2)$ phase boundary, the
specific heat peaks first decrease and then diverge again with
larger lattice sizes. Similar behavior was observed for the 4th
order cumulant: for the small lattice, there is more than one
diverging correlation length; but if the lattice sizes are big
enough, only one dominates. Therefore, in certain range of small
lattice sizes, the behavior is easy to confuse with XY-like\cite{kos73}, and
GPU accelerated simulations of large lattice sizes is essential. For
positive fields and low temperatures there is hysteresis in the m
vs. H data indicating the presence of first-order transitions, but
at higher temperature the data obtained for increasing and
decreasing fields are essentially identical.

\begin{figure}
\includegraphics[width=0.9\columnwidth]{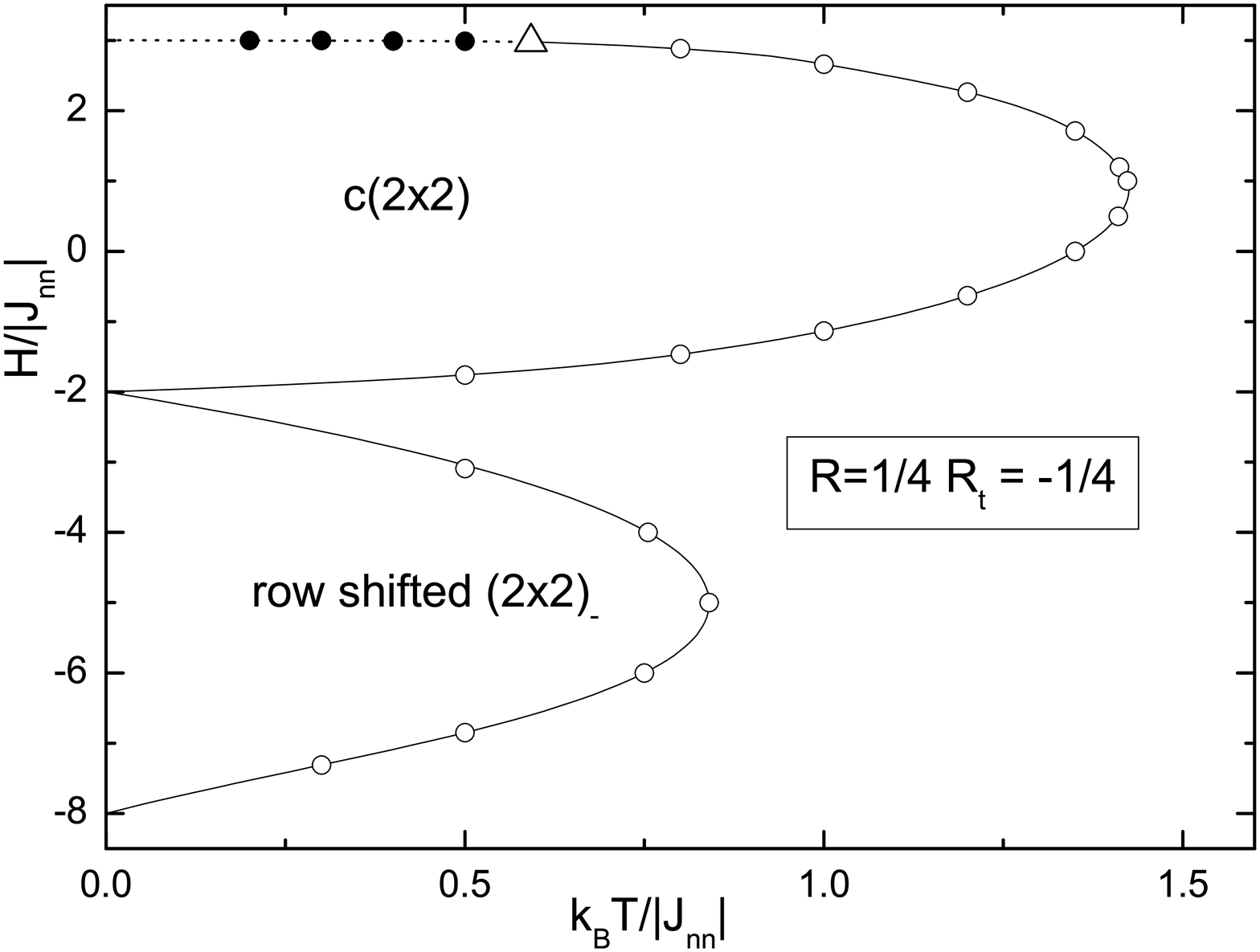}
\caption{Phase diagram in magnetic field-temperature space for $R = 1/4, R_t = -1/4$.
The solid curves are second-order phase boundaries and the dashed line
indicates first-order transitions. The open triangle indicates the location of a
tricritical point.} \label{f5}
\end{figure}

\begin{figure}
\includegraphics[width=0.9\columnwidth]{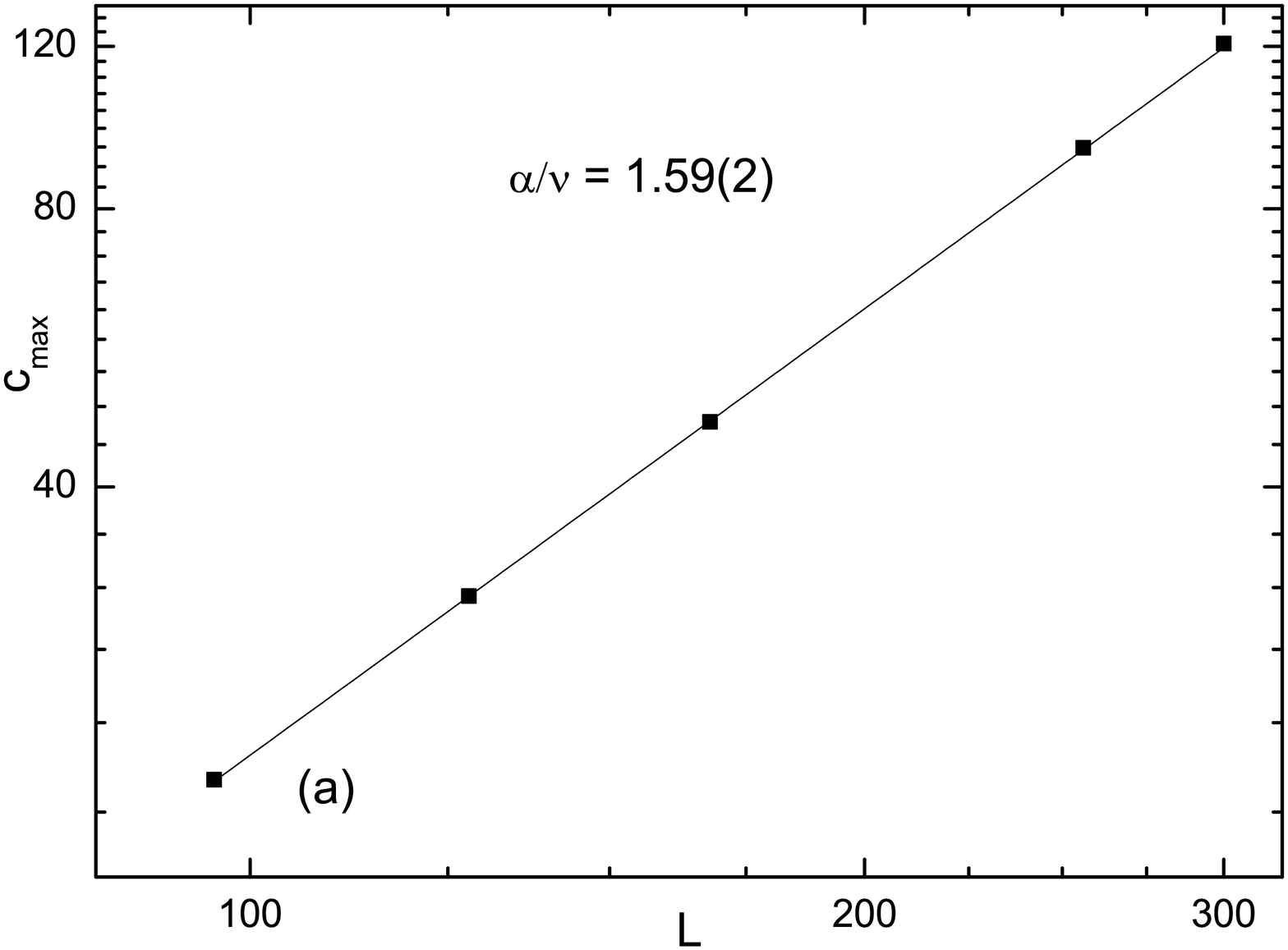}
\vspace{0.2cm}\\
\includegraphics[width=0.9\columnwidth]{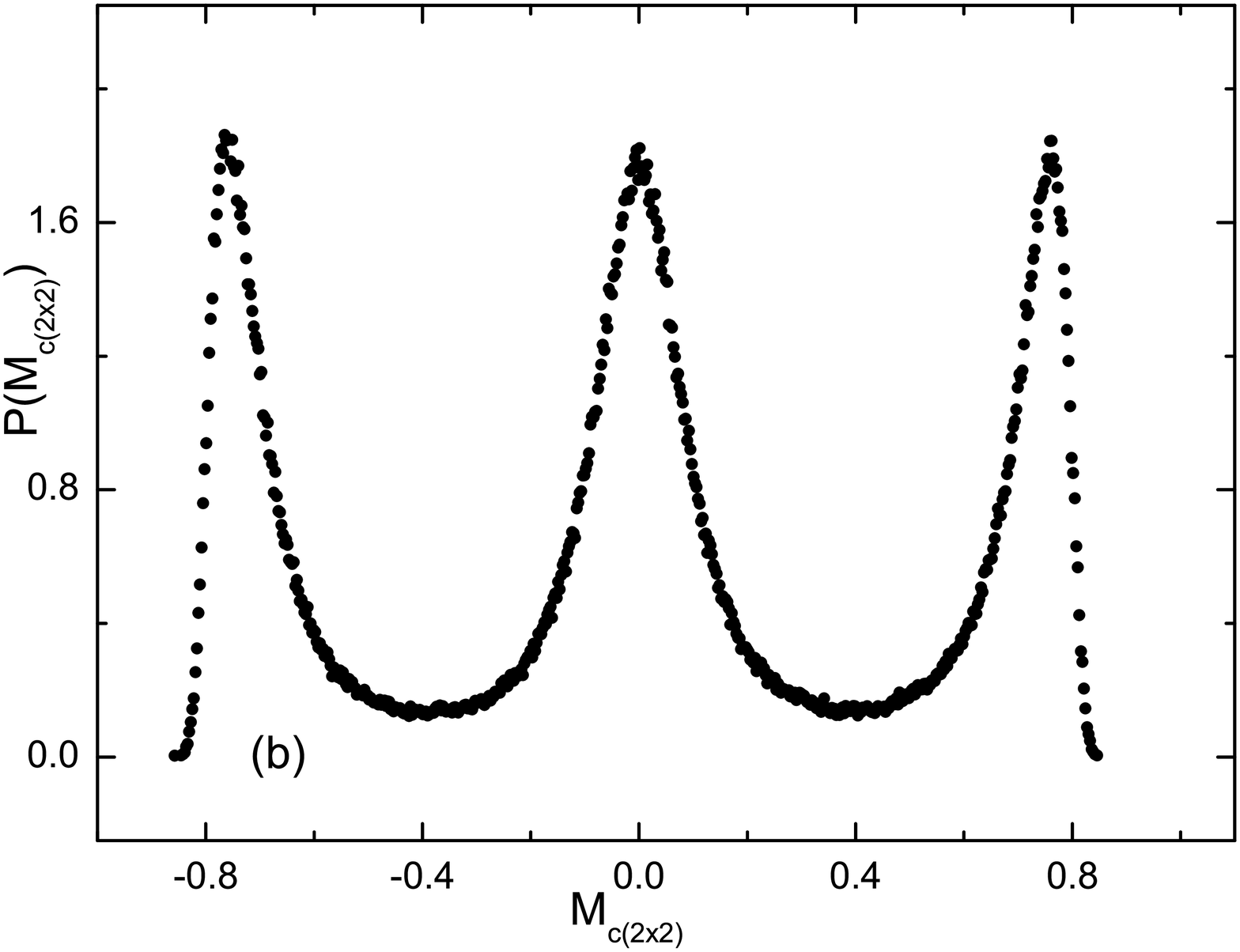}
\caption{Tricritical point(T=0.5915(4), H=2.98073(8)) for $R = 1/4, R_t = -1/4$: (a) Curve fit of the specific heat peaks.
(b) The density distribution of order parameter $M_{c(2\times2)}$ for $L=256$.} \label{fa1}
\end{figure}

The resultant phase diagram in field-termperature space is shown in
Fig.~\ref{f5}.
 The $c(2\times2)$ phase is separated from the
disordered phase on the high field side by a phase boundary which
contains a tricritical point but on the low field side the
transition appears to stay second-order down to the lowest
temperature studied, and the row-shifted $(2\times2)$ state is also
bounded by a line of second order transitions. Since the transition
from the $c(2\times2)$ ordered phase to the disordered phase should
belong to the Ising universality, we expect
$\alpha/\nu=0$(logarithm) for a second-order phase transition, and
$\alpha/\nu=2$ for a first-order phase transition. At the
tricritical point, the exact(conjectured) value for the exponent
$\alpha/\nu$ is $\frac{8}{5}$, which is supported by many
renormalization group calculations\cite{nij79}. Therefore, we
estimate the exponent from finite size behavior of specific heat
peaks, $c_{max}\sim L^{\alpha/\nu}$, near the connecting section of
the first- and second-order transition line, and found the
tricritical point is close to $k_BT/|J_{nn}|=0.592$. To confirm our
estimation and get more accurate location, we also calculate the
density distribution of the order parameter, as shown in
Fig.~\ref{fa1}. The final estimation of the tricirtial point is
$k_BT/|J_{nn}|=0.5915(4), H/|J_{nn}|=2.98073(8)$, and the evaluated
exponent $\alpha/\nu=1.59(2)$ agrees nicely with the predicted
value.

\begin{figure}
\includegraphics[width=0.9\columnwidth]{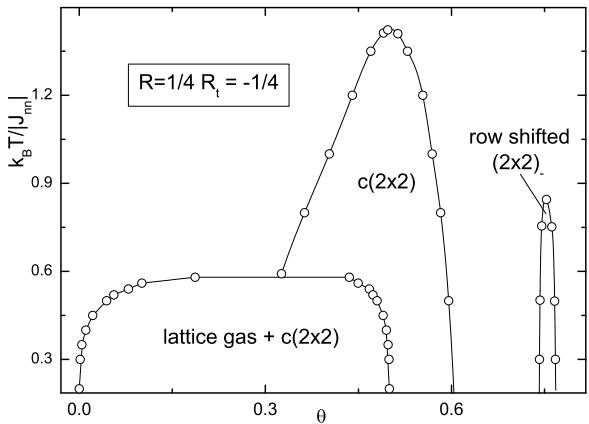}
\caption{Temperature-coverage phase diagram for $R = 1/4, R_t = -1/4$.} \label{f6}
\end{figure}

The corresponding phase diagram in coverage-temperature space is
shown in Fig.~\ref{f6}.
Here we see that the $c(2\times2)$ phase and the L.G.+$c(2\times2)$
coexistence phase, which is present below the tricritical point,
appear over substantial ranges of $\theta$ and $T$, whereas the
row-shifted $(2\times2)$ phase is actually confined to a very narrow
range of coverage.  This phase diagram is substantially different
from that predicted in Fig.~\ref{f2}c, and in particular there is no
triple point. However, If third-nearest- neighbor
two-body interactions are added, the ground state degeneracy for the
$(2\times2)$ state will be removed. Then, tricritical points involving the $(2\times 2)$ phase
 could occur, and triple points in the region of the first order transition perhaps as well, i.e., the predicted phase diagram in Fig.~\ref{f2}c could then be valid.

\begin{figure}
\includegraphics[width=0.9\columnwidth]{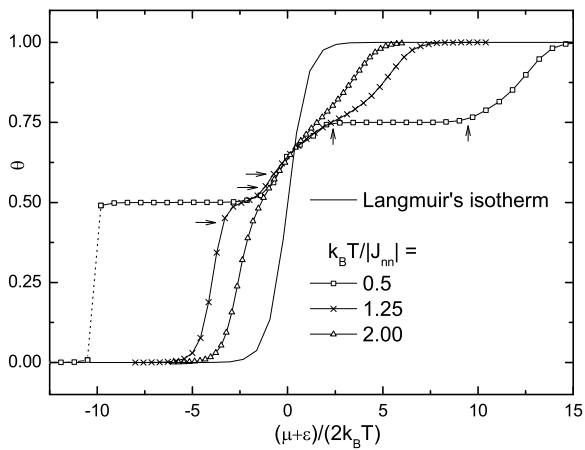}
\caption{Adsorption isotherms of the lattice gas model with $R = 1/4, R_t = -1/4$.
The arrows mark the second-order phase transitions.} \label{f7}
\end{figure}
In Fig.~\ref{f7} we show adsorption isotherms which are obtained for
several different temperatures and for comparison include the
Langmuir isotherm which would be correct for a non-interacting
lattice gas. The jump in the low temperature data shown by the
dotted line clearly locates the first-order transition; but the
second-order transitions, indicated by the arrows, are extremely
difficult to identify from the adsorption isotherms.  The step-like
behavior of the lowest temperature adsorption isotherm shown in this
figure is not dissimilar to the multiple risers which are seen for
multilayer adsorption, but here it merely represents multiple
transitions within a single layer!\\

B. ~~ $R=1/2, R_t = -1$

\begin{figure}
\includegraphics[width=0.9\columnwidth]{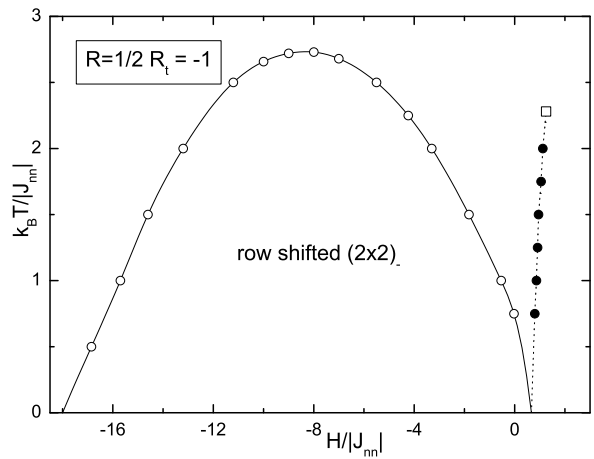}
\caption{Field-temperature phase diagram for $R = 1/2, R_t = -1$.  The solid
curves are second-order transitions and the dashed lines show first-order
phase boundaries. The open square indicates the location of the
terminating critical point.} \label{f8}
\end{figure}

The same thermodynamic properties were determined as were described
in Sec. A, and since there were no significant differences in the
nature of the results, we shall not show any raw data for this case.
The resultant phase diagram in H-T space is shown in Fig.~\ref{f8}.
A line of first-order transitions, terminating in a critical point,
separates a lattice liquid from a lattice gas state, and a line of
second-order transitions bounds a row-shifted $(2\times2)$ phase.

Since the lack of symmetry among the two different phases at the
critical point that terminates the first order line, the relevant
scaling fields $\tau, h$ are comprised by linear combinations of the
thermodynamic fields $T, H$ as\cite{reh73}

\begin{eqnarray}
\tau = T-T_c + s(H-H_c)
\end{eqnarray}
\begin{eqnarray}
h = H-H_c + r(T-T_c)
\end{eqnarray}
where s and r are parameters controlling the extent of field mixing.
As a result, the associated conjugate scaling operators ${\cal E},
{\cal M}$ are also linear combinations of the spin-spin interaction
energy density u and the magnetization m as
\begin{eqnarray}
{\cal E}=\frac{u-rm}{1-rs} \qquad \qquad \qquad \qquad \\
{\cal M}=\frac{m-su}{1-rs} \qquad \qquad \qquad \qquad \\
m=\frac{1}{N}\sum_{i} \sigma_i \qquad \qquad \qquad \quad \enspace \\
u=\frac{1}{N}(\sum_{i \neq j} \sigma_i \sigma_j +R \sum_{i \neq k} \sigma_i \sigma_k +R_t \sum_{i \neq j \neq k} \sigma_i \sigma_j \sigma_k)
\end{eqnarray}
where N is the total number of spins. According to the finite-size
scaling\cite{bru92}, the joint probability distribution $p_L({\cal
E}, {\cal M})$ near criticality should obey the following scaling
ansatz:

\begin{eqnarray}
p_L({\cal E}, {\cal M})\simeq\Lambda^+_{\cal M}\Lambda^+_{\cal E}\tilde{p}_{{\cal E}, {\cal M}}(\Lambda^+_{\cal M}\delta{\cal M},\Lambda^+_{\cal E}\delta{\cal E},\Lambda_{\cal M}h,\Lambda_{\cal E}\tau)
\end{eqnarray}
where,

\begin{eqnarray}
\Lambda_{\cal E}=a_{\cal E}L^{1/\nu},\quad \Lambda_{\cal M}=a_{\cal M}L^{d-\beta/\nu} \\
\Lambda_{\cal E}\Lambda^+_{\cal E}=\Lambda_{\cal M}\Lambda^+_{\cal M}=L^d \qquad \qquad\\
\delta{\cal M} = {\cal M} - <{\cal M}>_c,\quad \delta{\cal E} = {\cal E} - <{\cal E}>_c
\end{eqnarray}
The subscripts c denotes that the averages are taken at criticality.
For appropriate choices of the nonuniversal factors $a_{\cal E}$ and
$a_{\cal M}$, function $\tilde{p}_{{\cal E}, {\cal M}}$ would be
universal. After integration over ${\cal E}$, exactly at
criticality, where $h=\tau=0$, one has

\begin{eqnarray}
p_L({\cal M}) \simeq a_{\cal M}^{-1}L^{\beta/\nu}\tilde{p}^*_{\cal M}(L^{\beta/\nu} a_{\cal M}^{-1}\delta{\cal M})
\end{eqnarray}
where the function $\tilde{p}^*_{\cal M}$ characterizes the
universality class, the form of which has been well established for
the two-dimensional Ising model.
\begin{figure}
\includegraphics[width=0.9\columnwidth]{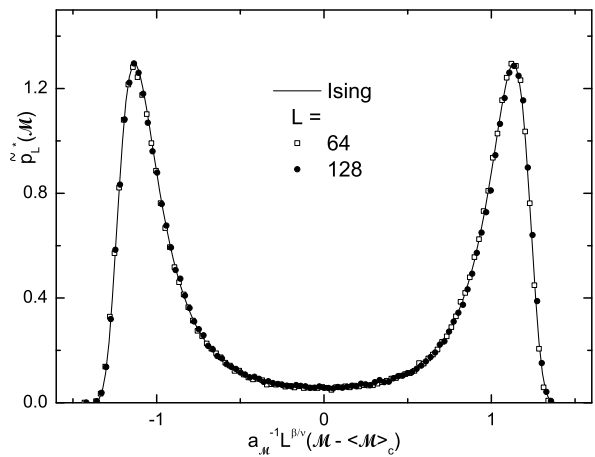}
\caption{The density distributions for $L=64$ and $128$ at the critical point $k_BT/|J_{nn}|=2.2738(4), H/|J_{nn}|=1.24925(13)$.
The full curve is the corresponding distribution for the two-dimensional Ising model for $L=400$. } \label{fa2}
\end{figure}
In Fig.~\ref{fa2}, we plot the density distribution function at
estimated criticality $k_BT/|J_{nn}|=2.2738(4),
H/|J_{nn}|=1.24925(13)$ with the controlling parameter $s=-0.30(2)$
for $L=64$ and $128$. The superimposed curve is the corresponding
distribution for the two-dimensional Ising model for $L=400$. The
nonuniversal factors $a_{\cal M}$ for each lattice sizes is chosen
in such a way that the variable $a_{\cal M}^{-1}L^{\beta/\nu}({\cal
M}-<{\cal M}>_c)$ has unit variance.

As for the critical exponents for the continuous transition from the
row-shifted $(2\times2)$ phase to the paramagnetic phase, the
correlation length exponent $\nu$ changes along the transition line,
but the reduced exponents $\gamma/\nu$ and $\beta/\nu$ seems to
belong to Ising universality, similar to what we found in Ref
\cite{yin09}.

\begin{figure}
\vspace{0.5cm}
\includegraphics[width=0.9\columnwidth]{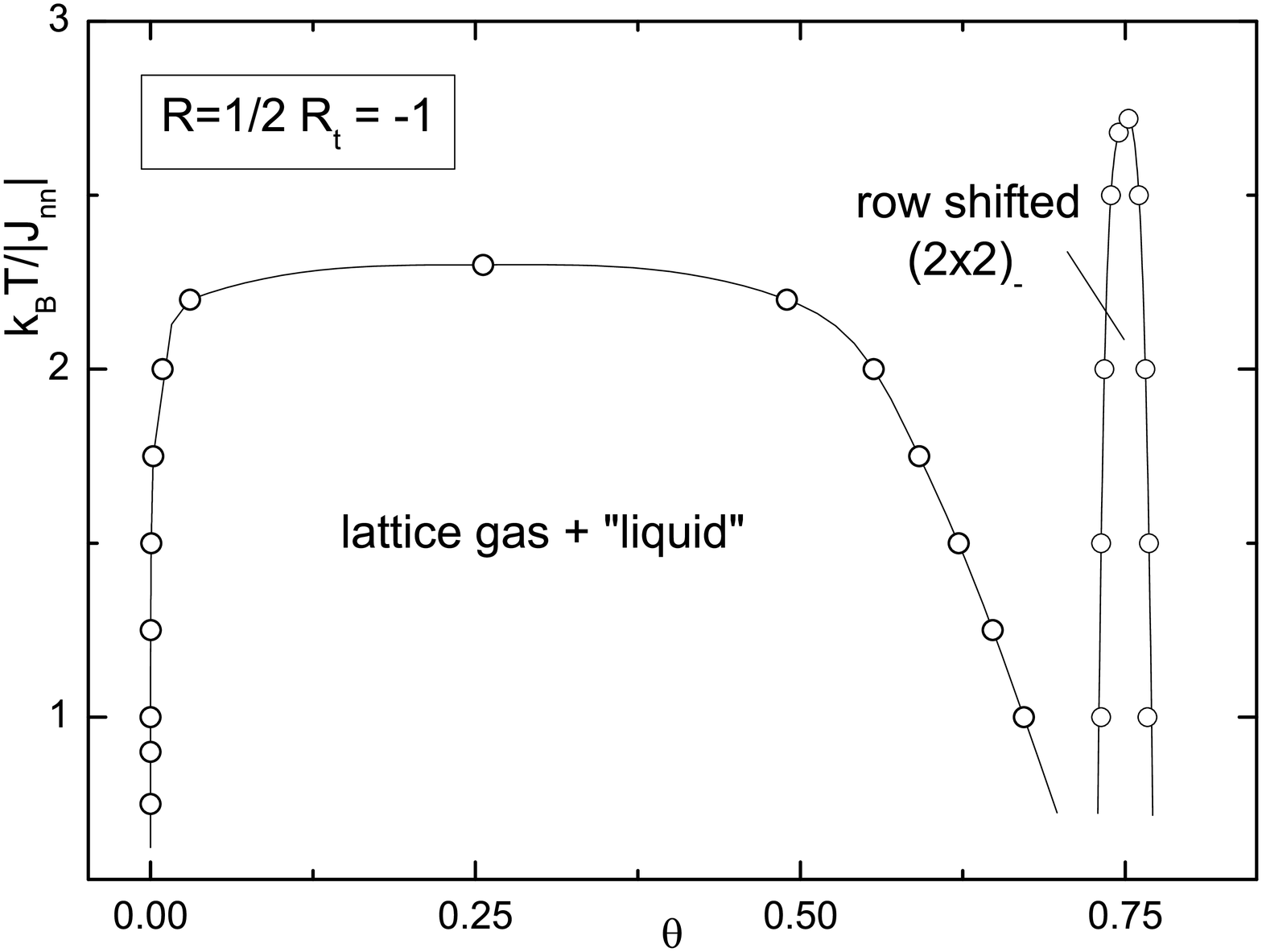}
\caption{Temperature-coverage phase diagram for $R = 1/2, R_t = -1$.} \label{f9}
\end{figure}

The phase diagram is replotted in coverage-temperature space in
Fig.~\ref{f9}.
Here, too, we see that the row-shifted $(2\times2)$ phase is present
only over a relatively narrow range of coverages, but the L.G. +
L.L. coexistence phase is stable over a much larger region of $T -
\theta$ space.  Comparing with Fig.~\ref{f2}d, we see that there are
qualitative differences between the actual behavior and the phase
diagrams which had previously been ``guessed''; but again, if the
degeneracy allowing for row-shifted structures were removed, the
predicted phase diagram in Fig.~\ref{f2}d might hold.

\section{Conclusions}

Monte Carlo simulations have been used to extract phase diagrams for
simple models on a square lattice with three-body interactions which
are larger in magnitude than those which have been previously
studied.  We find qualitatively different behavior than that which
had been suggested in Ref ~\cite{bin81}. If third-nearest- neighbor
two-body interactions are added, the ground state degeneracy for the
$(2\times2)$ state will be removed; but of course there is no
guarantee that the finite temperature behavior will not show
remnants of this effect.  These results further demonstrate the
complexity which may be found in relatively simple models with
competing interactions. This problem is of interest to statistical
mechanics in its own right and we believe that further Monte Carlo
studies of such models will continue to display many of the features
observed in experimental studies of adsorbed monolayers.

\section{Acknowledgements}

We wish to thank T. L. Einstein and K. Binder for illuminating comments and
suggestions, J. L. Li and C. W. Chen for sharing preliminary data,
and S.-H. Tsai for providing the critical density distribution of
the 2D Ising model for $L=400$. This research was supported in part
by National Science Foundation grant DMR-0810223.


\end{document}